\documentclass[conference]{IEEEtran}
\IEEEoverridecommandlockouts

\usepackage{array}
\newcommand{\PreserveBackslash}[1]{\let\temp=\\#1\let\\=\temp}
\newcolumntype{C}[1]{>{\PreserveBackslash\centering}p{#1}}
\usepackage{multirow}
\usepackage{graphicx}
\usepackage{textcomp}
\usepackage{amsmath}

\usepackage[colorlinks = true,
            linkcolor = blue,
            urlcolor  = blue,
            citecolor = blue,
            anchorcolor = blue]{hyperref}

\usepackage{bbm}
\newcommand{\minus}{\scalebox{1.0}[1.0]{$-$}}
\usepackage{dsfont}
\usepackage{color}

\usepackage{booktabs,dcolumn}
\usepackage{tabularx} 

\usepackage{array}
\newcolumntype{Z}{>{\centering\let\newline\\\arraybackslash\hspace{0pt}}X}

\usepackage{algorithm}
\usepackage{algorithmic}

\usepackage{cite}
\usepackage{xcolor}
\usepackage{amsmath,amssymb,amsfonts}
\usepackage{float}

\newsavebox{\foobox}
\def\BibTeX{{\rm B\kern-.05em{\sc i\kern-.025em b}\kern-.08em
    T\kern-.1667em\lower.7ex\hbox{E}\kern-.125emX}}
\begin{document}

\title{Nothing Stands Alone: Relational Fake News Detection with Hypergraph Neural Networks}

\author{\IEEEauthorblockN{Ujun Jeong\IEEEauthorrefmark{1},
Kaize Ding\IEEEauthorrefmark{1}, Lu Cheng\IEEEauthorrefmark{2},
Ruocheng Guo\IEEEauthorrefmark{3}, Kai Shu\IEEEauthorrefmark{4} and Huan Liu\IEEEauthorrefmark{1} }
\IEEEauthorblockA{\IEEEauthorrefmark{1}School of Computing and Augmented Intelligence, Arizona State University\\
\IEEEauthorrefmark{2}Department of Computer Science, University of Illinois Chicago\\
\IEEEauthorrefmark{3}Bytedance AI Lab, London\\
\IEEEauthorrefmark{4}Department of Computer Science, Illinois Institute of Technology\\
Email: \IEEEauthorrefmark{1}\{ujeong1, kding9, huanliu\}@asu.edu,
\IEEEauthorrefmark{2}lucheng@uic.edu,
\IEEEauthorrefmark{3}rguo.asu@gmail.com,
\IEEEauthorrefmark{4}kshu@iit.edu}}

\maketitle

\begin{abstract} Nowadays, fake news easily propagates through online social networks and becomes a grand threat to individuals and society. Assessing the authenticity of news is challenging due to its elaborately fabricated contents, making it difficult to obtain large-scale annotations for fake news data. Due to such data scarcity issues, detecting fake news tends to fail and overfit in the supervised setting. Recently, graph neural networks (GNNs) have been adopted to leverage the richer relational information among both labeled and unlabeled instances. Despite their promising results, they are inherently focused on pairwise relations between news, which can limit the expressive power for capturing fake news that spreads in a group-level. For example, detecting fake news can be more effective when we better understand relations between news pieces shared among susceptible users. To address those issues, we propose to leverage a hypergraph to represent group-wise interaction among news, while focusing on important news relations with its dual-level attention mechanism. Experiments based on two benchmark datasets show that our approach yields remarkable performance and maintains the high performance even with a small subset of labeled news data.
\end{abstract}

\section{Introduction}
With the growing amount of information that spreads through online social networks, people are at high risk of being exposed to fake news~\cite{vosoughi2018spread}. Online social networks often facilitate the spread of fake news as people tend to search for, recall, and share information that supports one’s prior beliefs due to \textit{confirmation bias}~\cite{nickerson1998confirmation}. Further, the \textit{echo chamber effect}\cite{cinelli2021echo} reinforces
this prism of reality among people who share the same opinion. Consequently, fake news jeopardizes the public trust in government and accelerates polarization in society~\cite{shu2020early}.

Nevertheless, assessing the authenticity of news is difficult, as fake news is written to intentionally mislead people with elaborately manipulated facts and narratives. The contents and topics covered by fake news are also diverse, which makes it extremely laborious to obtain large amounts of annotated fake news. Existing works in fake news detection mainly focus on the supervised setting where abundant labeled data is given~\cite{shu2019beyond, monti2019fake, ren2021fake}. In this case, the detection models would easily overfit and fail to make accurate predictions when only limited labeled data is available. Therefore, how to solve the problem of fake news detection under the low-resource scenario has become a challenging task that requires urgent research efforts.

\begin{figure}
\centering
\vspace{-0.5cm}
\includegraphics[width=0.5\textwidth]{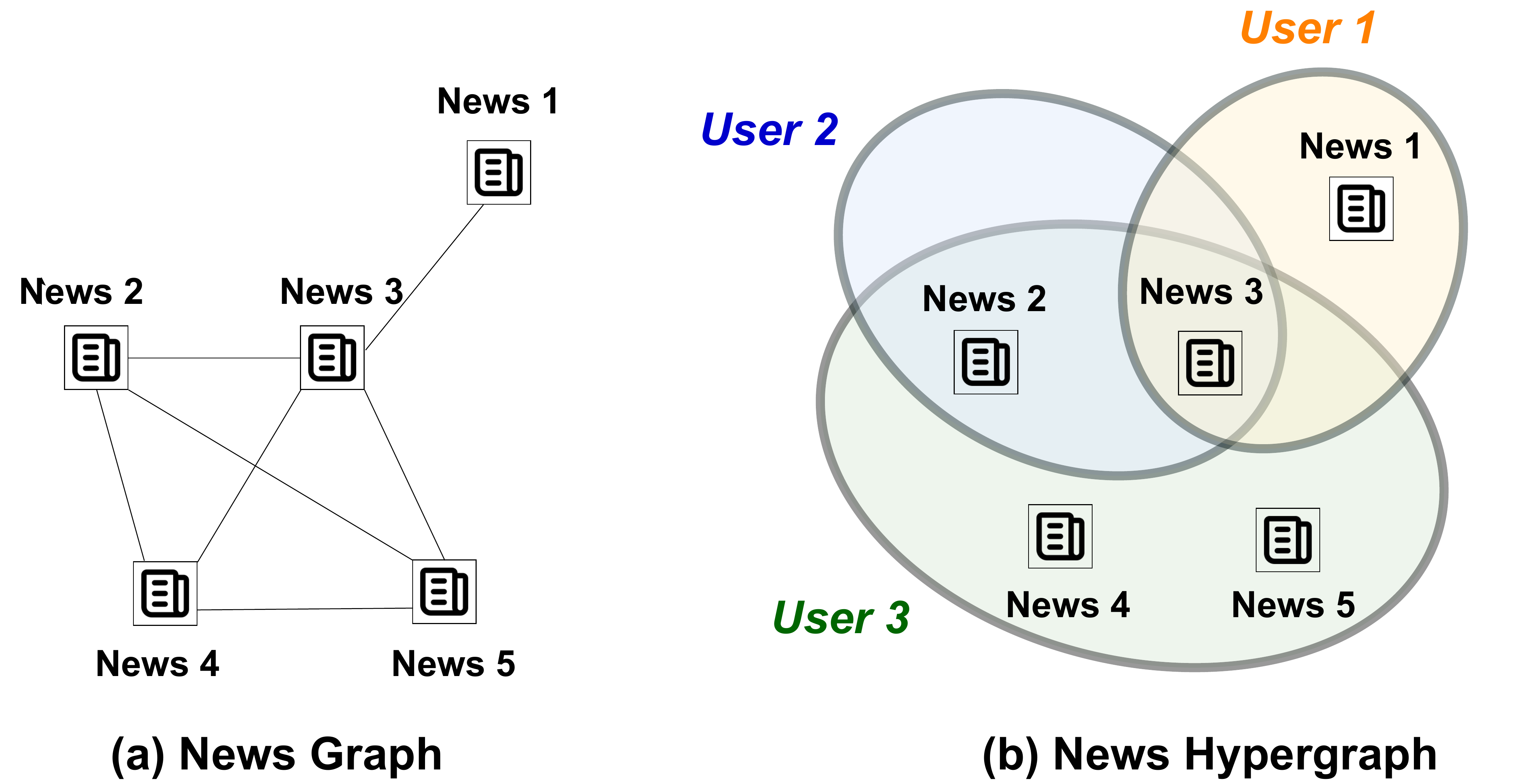}
\caption{Examples of a news graph (a) and a news hypergraph (b) representing users' news sharing behavior on a social network. Each node denotes a news piece, and nodes are connected when news are shared by the same user.
} \label {NewsGraphHypergraph}

\vspace{-0.5cm}
\end{figure}

In this paper, we aim to investigate whether capturing meaningful interactions between news can improve the performance of fake news detection under the low-resource setting. However, it is a non-trivial task due to the following reasons: (1) Though graph neural networks (GNNs) have been applied to fake news detection~\cite{nguyen2020fang, ren2021fake} to capture the relational information, this line of work assumes that the news interactions are based on pairwise relations, which may omit the important group-wise interactions among different news pieces. In fact, many fake news pieces spring up organically~\cite{bernardi2012narrative}, that being said, fake news is often shared by a group of similar users. This means that fake news can be better understood by capturing such group-wise interactions; (2) Most graph-based fake news detection methods merely consider the relations for the labeled news pieces that are naturally formed during the proliferation process of news on a social network~\cite{dou2021user, han2020graph, bian2020rumor}. However, the information from unlabeled news pieces and their interactions with labeled news are largely ignored. Hence, how to incorporate the unlabeled data that can potentially benefit the propagation of label information is another key challenge to solve.

To address the aforementioned challenges, we first propose to use a \textit{hypergraph} as the modeling basis. A hypergraph can naturally represent group-wise interactions by connecting multiple ($\geq 2$) nodes to its special edge called a \textit{hyperedge}. For example, Fig.~\ref{NewsGraphHypergraph} shows users' news sharing behavior on a social network. The news hypergraph (Fig.~\ref{NewsGraphHypergraph}(b)) can fuse the crucial group-wise interaction that \textit{News 3}, \textit{News 4}, and \textit{News 5} are shared by the same user. In contrast, the news graph (Fig.~\ref{NewsGraphHypergraph}(a)) only captures the pairwise relations using the following edges: (\textit{News 3}, \textit{News 4}), (\textit{News 3}, \textit{News 5}), and (\textit{News 4}, \textit{News 5}), which can lead to the misinterpretation that the three news pieces are shared by three different users. Moreover, we design the problem of fake news detection as a semi-supervised node classification task in the hypergraph, where nodes denote news pieces and hyperedges denote any group-wise interactions. This way we are able to leverage both labeled and unlabeled news based on different hyperedges and propagate the feature patterns of labeled news in the news hypergraph. To further detect fake news on the constructed news hypergraph, we propose a new hypergraph neural network model -- HGFND (\textbf{H}yper\textbf{G}raph for \textbf{F}ake \textbf{N}ews \textbf{D}etection). Considering the fact that noisy and task-irrelevant relations in the hypergraph (e.g., news shared mistakenly~\cite{zhou2022fake}) may jeopardize the learning process, HGFND adopts a dual-level attention mechanism to highlight useful news relations by measuring their importance in both node-level and hyperedge-level. Experiments on two real-world benchmark datasets for fake news detection (i.e., Politifact, Gossipcop) show that the proposed model outperforms the state-of-the-art methods by a large margin with both full and limited number of labeled data. Our main contributions are summarized as follows:

\begin{itemize}
     \item To the best of our knowledge, this is the first work that incorporates group-wise interactions among news pieces by leveraging a hypergraph for fake news detection. 
	\item We propose to construct the hypergraph based on social context and news content. We also define three types of hyperedges based on user, time, and entity shared in news.
	\item We propose a novel hypergraph neural network, which outperforms graph neural networks for fake news detection with both full and limited number of labeled data.
    \item Our model can jointly capture propagation patterns of news while leveraging the  information from unlabeled news and their relations with labeled news.

\end{itemize}
\section{Related Work}
In this chapter, we focus on introducing various methods used in fake news detection and reviewing a series of studies on graph neural networks in relational machine learning.

\subsection{Fake News Detection}
The most basic characteristic of news is its textual content. Early researches about fake news detection mostly relied on the content in the body or headline of news to understand its linguistic features and compare it with relevant evidences retrieved from Wikipedia\cite{hanselowski2018retrospective, thorne2018fever}. For example, a series of work has shown promising result in encoding the news textual information such as Riedel et al.~\cite{riedel2017simple} and other BERT-based models~\cite{kaliyar2021fakebert}. On the other hand, social context-based approaches include user-driven social engagements of news consumption on social network platforms. For example, CSI~\cite{ruchansky2017csi} incorporates the behavior of both parties, users and articles, and the group behavior of users who spread news. dEFEND~\cite{shu2019defend} utilizes user and news comments based on hierarchical attention network~\cite{yang2016hierarchical}. Han et al.~\cite{han2020graph} utilize user features such as the number of followers and friends without any textual information related to news to detect. Cheng et al.~\cite{cheng2021causal} show causal relationships between user's attributes and susceptibility of users. A series of works using GCN~\cite{kipf2016semi} have been used for modeling news propagation tree such as GCNFN~\cite{monti2019fake}, BiGCN~\cite{bian2020rumor} and GNN-CL~\cite{han2020graph}. UPFD~\cite{dou2021user} benchmarks FakenewsNet~\cite{shu2020fakenewsnet} by jointly capturing the news contents and surrounding exogenous context of news through propagation tree.  GTN~\cite{matsumoto2021propagation} and TGNF~\cite{song2021temporally} consider time difference between tweets/retweets in propagation trees using time-aware message aggregation and attention networks. Silva et al.~\cite{silva2021propagation2vec} focus on predicting propagation patterns of fake news in its early stage by reconstructing news propagation trees.   Various methods have been applied to leverage auxiliary information related to news. HGAT~\cite{ren2021fake} leverages a heterogeneous graph to model pairwise relations between news, creators, and subjects through node and semantic-level attention mechanism. A hypergraph has recently drawn attention in fake news detection by clustering the news content sharing patterns of users~\cite{muhuri2021hypergraph}, or jointly modeling topic relations between news~\cite{borse2022fake} with hierarchical attention network.



\subsection{Graph Neural Networks}
Graph Neural Network (GNN) models aim to obtain better node representations via message passing among local neighbors in the graph using neighborhood aggregation~\cite{ding2020graph,ding2022data}. GCN~\cite{kipf2016semi} can be explained as a mean-pooling neighborhood aggregation. GraphSAGE~\cite{hamilton2017inductive} uses node features with max pooling or LSTM based aggregation, which can be used in inductive representation learning. GAT~\cite{velivckovic2017graph} incorporates trainable attention weights to learn weights on neighbors when aggregating neighboring information about a node. Furthermore, a heterogeneous graph has been adopted to represent various types of nodes and  edges~\cite{wang2019heterogeneous} to enhance expressive power of the graph. However, existing graph neural networks only consider pairwise relationships between objects, and cannot apply to non-pairwise relation learning. Recently, hypergraph-based methods have drawn much attention due to its expressive power to represent high-order relations. Convolutional hypergraph neural networks~\cite{bai2021hypergraph} has shown promising results by applying the convolution operation on graph to hypergraph in semi-supervised learning setting. For instance, HGNN~\cite{feng2019hypergraph} proposes to build hyperedges by connecting nodes that are semantically similar based on a distance metric. While convolutional hypergraph neural networks expand the convolution operation on a graph to a hypergraph, hypergraph attention further enhances the capacity of representation learning by leveraging an attention module~\cite{bai2021hypergraph}. For example, HyperGAT~\cite{ding2020more} uses dual-level attention mechanism to capture high-order information between words and sentences for document classification in an inductive setting. Additionally, hypergraph neural networks have been widely applied in many other applications, such as recommendation system~\cite{wang2020next, wang2021session}, visual object detection~\cite{feng2019hypergraph}, and advertisement classification~\cite{jeong2022classifying}.


Compared to past approaches in fake news detection, our approach has a clear distinction in that it can represent group-wise interactions among news with hypergraph neural networks, which leverages the hypergraph constructed from both social context and news content. The proposed model helps to identify important group-wise interactions for fake news detection with dual-level attention mechanism for nodes (news) and hyperedges in the stage of message aggregation.


\section{Preliminaries}
In this section, we introduce the problem of relational fake news detection. To this end, we compare the definition of a propagation tree and a hypergraph and define the problem statement based on these structures. Foremost, the formal definition of a propagation tree is as follows:

\hfill

\paragraph{\textbf{Propagation Tree}}
Given total $\mathit{N}$ number of news as $\mathcal{X} = \{x_i\}^\mathit{N}_{i=1}$ and user engagement associated to news $x_i$ as $\mathcal{U}_i = \{u_{t_j}\}^\mathit{|\mathcal{U}|}_{j=1}$ where $t_j$ denotes the sequence of timestamp of a user's engagement such as tweet and retweet. A propagation tree is defined as $P_i = ( \{x_i\} \bigcup \mathcal{U}_i , \mathbf{A} )$ where $\mathbf{A}$ is an adjacency matrix estimated by time-inferred diffusion~\cite{vosoughi2018spread}.

\hfill

A propagation tree is designed to represent the cascade of news that spreads on a social network through user engagements. While a propagation tree is used for a single news piece, we use a hypergraph to represent the relations among multiple news pieces. The formal definition of a hypergraph is as follows:

\hfill

\paragraph{\textbf{Hypergraph}}
A hypergraph is defined as $\mathcal{G} = (\mathcal{V}, \mathcal{E})$, where $\mathcal{V} = \{v_i\}^\mathit{N}_{i=1}$ denotes a set of nodes in the graph, and $\mathcal{E} =  \{e_j\}^\mathit{M}_{j=1}$ represents a set of hyperedges. Different from a graph, a hypergraph can connect two or more nodes by a hyperedge. A hypergraph $\mathcal{G}$ can be represented by an incidence matrix $\mathbf{H} \in \mathds{R}^{\textit{N} \times \textit{M}}$, with entries defined as:

\begin{equation}
\mathbf{H}_{i,j} = 
    \begin{cases}
      1 & \text{if $v_{i} \in e_{j}$}\\
      0 & \text{if $v_{i} \notin e_{j}$}\\
    \end{cases}
\end{equation}

\hfill

The existing approaches to detecting fake news mainly focus on either to learn the characteristics of fake news individually and independently~\cite{riedel2017simple, shu2019defend} or use a graph to represent news interactions as pairwise relations~\cite{monti2019fake, ren2021fake}. However, a graph can be limited in representing the group-wise interactions when the relations between news pieces are higher-order~\cite{zhou2006learning}. To bridge this gap, we propose to define the problem of relational fake news detection as follows:

\begin{figure}
\centering
\includegraphics[width=0.35\textwidth]{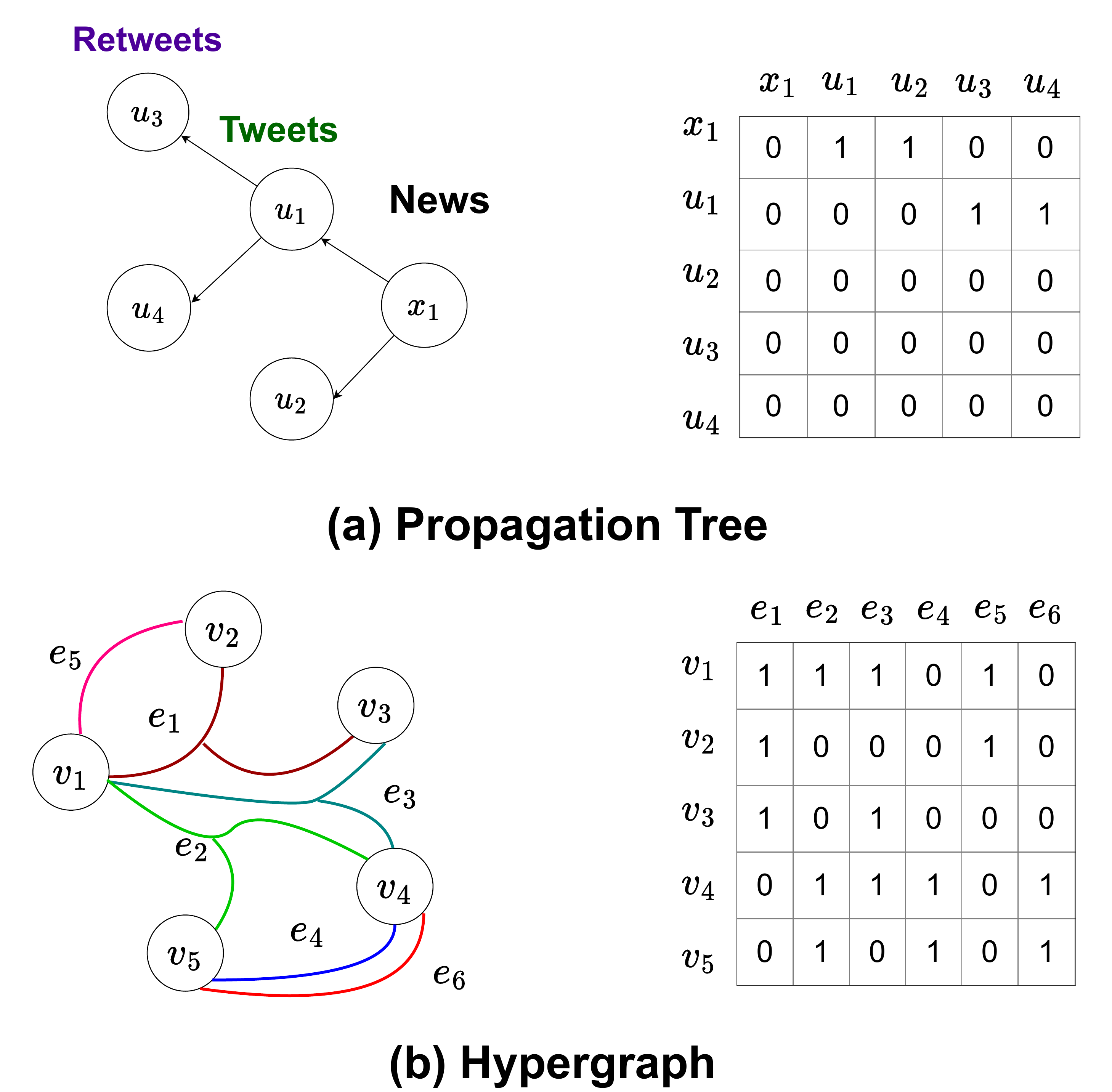}
\caption{A comparison between (a) a propagation tree and (b) a hypergraph} \label{ComparePropagationHypergraph}
\end{figure}
\hfill

\paragraph{\textbf{Problem Statement}} Given total \textit{N} number of news, the goal of relational fake news detection is to predict the labels of news into two labels $\mathcal{Y} = \{y\}^\mathit{N}_{i=1} \subset \{0, 1\}^\mathit{N}$ where news corresponds to a node in the hypergraph $\mathcal{G} = \{ \mathcal{V}, \mathcal{E} \}$. The hypergraph connects nodes (news) through a hyperedge based on the news contents $\mathcal{X} = \{x_i\}^\mathit{N}_{i=1}$ and their social context obtained from the propagation trees $\mathcal{P} = \{P_i\}^\mathit{N}_{i=1}$.

\section{Methodology}
\subsection{Constructing Hypergraph for News Relations}
\label{Hypergraph_Construction}
To construct the hypergraph for news relations, we assume there is a dataset that has already collected news content and user engagements, such as FakeNewsNet~\cite{shu2020fakenewsnet}. Based on this, we propose a hypergraph construction method for fake news detection using three types of hyperedges, as follows:
\hfill\\

\subsubsection{\textbf{User} (Shared User ID between News Pieces)} This type of hyperedge implies the news sharing behavior of a user. This means a hyperedge corresponds to a \textit{User ID}. To this end, we retrieve the unique \textit{User ID}s from tweet/retweet in the propagation tree of news. In order to Then, if a piece of news shares the same \textit{User ID} as other news pieces, we connect them to a hyperedge.
\hfill\\

\subsubsection{\textbf{Time} (News Shared in Proximal Time Range)}
This type of hyperedge assumes that similar news can emerge and be shared in the same time. This means a hyperedge corresponds to the creation timestamp of tweet/retweet in the propagation tree of news. Since the timestamp is in seconds, there may not be many news shared at the exact same time. To solve this problem, we round the timestamp to the nearest day or hour when connecting news. The rounding method can be opted flexibly based on the performance on the task.
\hfill\\

\subsubsection{\textbf{Entity} (Shared Entity between News Contents)} This type of hyperedge aims to connect news that deal with similar contents by checking the entity shared between news pieces. For example, we connect news by a hyperedge if they share the entity `COVID-19'. To extract entities from news content, we employ an entity recognition tool from spacy~\cite{honnibal2017spacy}. Here, we adopt seven types of entities provided in spacy: organizations \textit{(ORG)}, people \textit{(PER)}, products \textit{(PROC)}, title of arts such as books and songs \textit{(WORK\_OF\_ART)}, nationalities or religious or political groups \textit{(NORP)}, non-GPE locations \textit{(LOC)}, and named events such as hurricanes, wars, and sports \textit{(EVENT)}.

\hfill

To construct the proposed hypergraph, we simply combine the incidence matrix of $\mathbf{H}^{user},\:\mathbf{H}^{time}$, and $\mathbf{H}^{entity}$ from three construction methods by $\mathbf{H} = \mathbf{H}^{user} \oplus \mathbf{H}^{time} \oplus \mathbf{H}^{entity}$ where $\oplus$ denotes concatenation operation. Kindly note that the hypergraph needs to be constructed before training the model.

\subsection{Leveraging News Relations for Fake News Detection}
\label{hypergraph attention network}
\begin{figure*}
\centering
\includegraphics[width=0.9\textwidth]{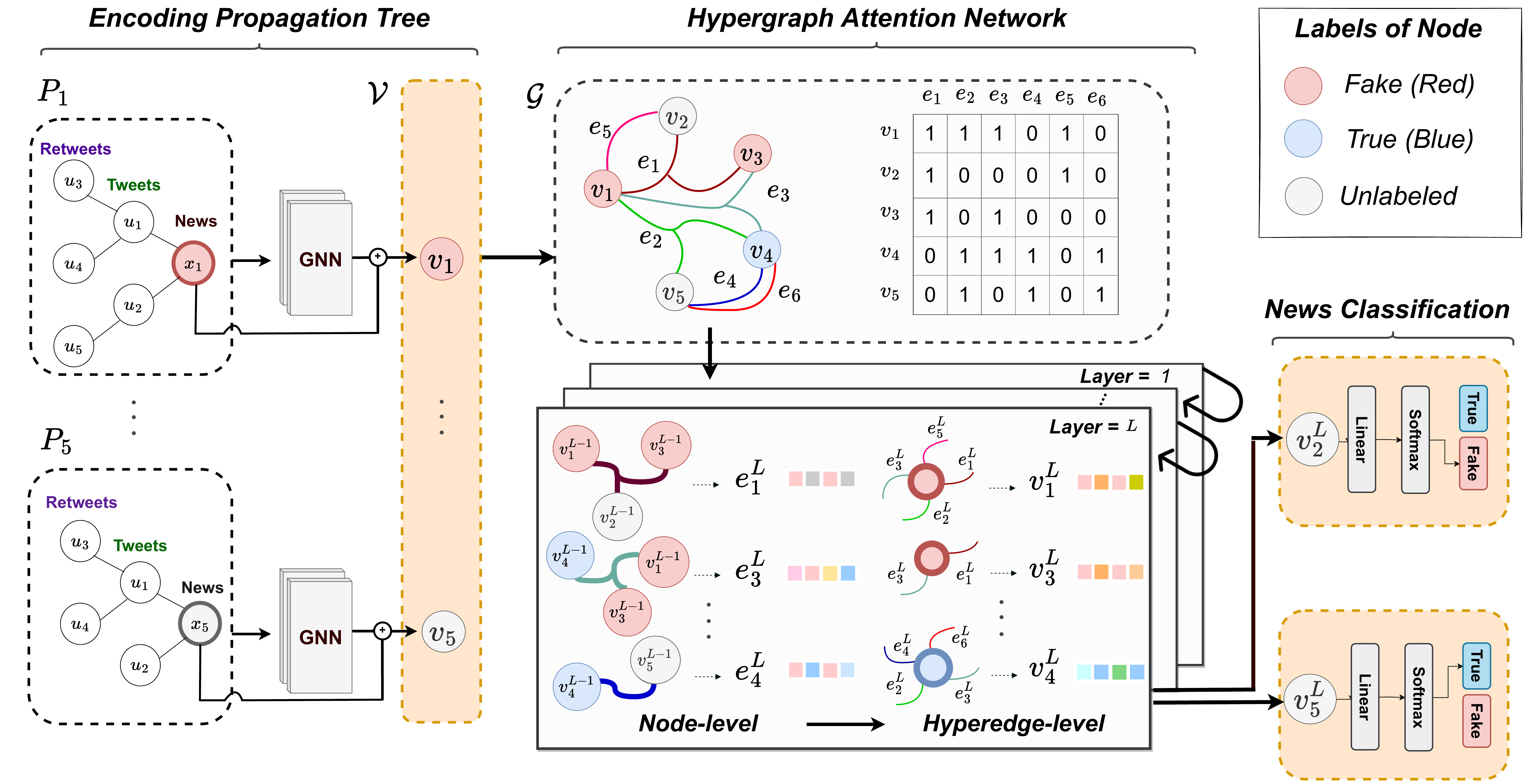}
\caption{Illustration of the proposed framework, which jointly models propagation trees and the hypergraph for classifying unlabeled news. The figure shows an example of transdcutive node classification where the number of news is 5 and the number of hyperedge is 6. The color of node in the hypergraph represents the label of news. i.e., fake (red), true (blue), and unlabeled (gray). The figure is best shown in color.} 
\label{proposed_model}
\end{figure*}

In this section, we propose our hypergraph neural network, well-designed for fake news detection. Fig.~\ref{proposed_model} shows how the proposed method can learn important correlations in the hypergraph while jointly modeling the propagation patterns of news. In the following, we explain about three major modules: (1) encoding  propagation tree for node representation in the hypergraph, (2) learning relational information between nodes and hyperedge using dual-level attention mechanism~\cite{ding2020more}, and (3) the final layer for news classification.

\begin{table}
\centering
\caption{Notations and Descriptions.}
\label{notations}
\begin{tabular}{cc}
\hline
 \textbf{Notation} & \textbf{Description}   \\
\hline
\hline
$\mathcal{X}$ & set of features for news contents\\
$\mathcal{U}_i$ & set of features for user engagements on news $x_i$  \\
$\mathcal{P}$ & set of propagation trees for $\mathcal{X}$ \\
$\mathcal{V}$ & set of nodes in the hypergraph  \\
$\mathcal{E}$ & set of hyperedges in the hypergraph  \\
$\overline{x}_i$ & representation of the propagation on news $x_i$  \\
$v^l_i$, $e^l_j$ & representation of $v_i$ and $e_j$  in the $\textit{l}$-th layer  \\
$\alpha_{i}$, $\beta_{j}$ & attention coefficient for $v_i$ and  $e_j$ \\
$h_i$, $r_j$  & hidden representation for $v_i$ and $e_j$   \\
$a_1$, $a_2$ & context vector for $h_i$ and  $r_j$ \\
$y_i, \hat{y}_i$ & ground-truth label and predicted logit of $v_i$  \\
\hline
\end{tabular}
\end{table}

\subsubsection{Encoding Propagation Tree}
Given a set of propagation trees $\mathcal{P}=\{P_i\}_{i=1}^{\textit{N}}$, we use Graph Neural Networks (GNNs) to  encode each propagation tree. To obtain the propagation pattern of news, we extract the root representation of the propagation tree $\overline{x}_i \in \mathbb{R}^{\mathit{F}}$ where $\mathit{F}$ is the size of the input feature matrix. Next, we concatenate  $\overline{x}_i$ with the original news features $x_i \in \mathbb{R}^{\mathit{F}} $ by skip-connection to generate a news representation enhanced with its propagation pattern. Then, we project it into $v^0_i \in \mathbb{R}^{d}$ for the initial node representation of the hypergraph with hidden dimension size $\mathit{d}$ as follows: 

\begin{equation}
\begin{aligned}
\overline{x}_i =  \textsc{Root}\left(GNN\left( P_i \right)\right),\\
    v^{0}_{i} = \mathit{f} \left(\sigma \left( x_i \oplus\overline{x}_i \right)\right),
\end{aligned}
\end{equation}
where $\sigma$ is the nonlinearity such as ReLU, $\textit{f}$ is a fully connected layer mapping $\mathbb{R}^{2 \times  \mathit{d}} \rightarrow \mathbb{R}^{ \mathit{d}}$, and GNN denotes a multi-layer graph neural network encoder. Specifically, we use GraphSAGE~\cite{hamilton2017inductive} as the backbone of the GNN module.

\subsubsection{Node-level Attention for Hyperedge Representation}
Given all the node representation $v^{l-1} \in \mathbb{R}^{|\mathcal{V}|\times \mathit{d}}$ in the layer $\textit{l}-1$, the first layer of the hypergraph attention network learns the representations of hyperedges $\mathcal{E}$. To reflect the different importance of nodes to hyperedge representation, we use a node-level attention mechanism to highlight those nodes that are important to the meaning of the hyperedge. Formally, nodes are aggregated to compute the hyperedge representation $e^{l}_j \in \mathbb{R}^{\mathit{d}}$ in the $\textit{l}$-th layer as follows:

    

\begin{equation}
    e^{l}_{j} = \sigma \left(  \sum_{v_{k} \in e_{j}} \alpha_{jk} \mathrm{W_{1}} v^{l-1}_{k} \right),
\end{equation}
where $v_{k}$ is a node connected to the hyperedge $e_{j}\in\mathcal{E}$, $\mathrm{W_1} \in \mathbb{R}^{ \mathit{d} \times  \mathit{d}}$ is a trainable weight matrix, and $a_1 \in \mathbb{R}^{|\mathcal{E}| \times |\mathcal{V}|}$ is a context vector. The attention coefficient $\alpha_{jk}$ for node representation $v^{l-1}_k$ in the $\textit{l}$-th layer is computed as follows:
\begin{equation}
\begin{aligned}
    \alpha_{jk} &= \frac{exp \left( a^{T}_{1} h_{k} \right)}{\sum_{v_{p} \in e_{j}} exp \left( a^{T}_{1} h_{p}\right)},\\
    h_{k} &= LeakyReLU \left(  \mathrm{W_{1}} v^{l-1}_{k}  \right),
\end{aligned}
\end{equation}

\subsubsection{Hyperedge-level Attention for Node Representation}
With all the hyperedge representations $e^{l} \in \mathbb{R}^{|\mathcal{E}|\times{ \mathit{d}}}$ in the $\mathit{l}$-th layer, we apply an edge-level attention mechanism to focus on the useful hyperedges for learning the next-layer representation of node $v_{i}$. This process can be expressed as:
\begin{equation}
    v^{l}_{i} = \sigma \left(  \sum_{e_{j} \in \mathcal{E}_{i}} \beta_{ij} \mathrm{W_{2}} e^{l}_{j} \right),
\end{equation}
where $v^{l}_{i}$ is the representation of node $v_{i}$ in the $\mathit{l}$-th layer, $\mathrm{W_2} \in \mathbb{R}^{\mathit{d} \times \mathit{d}}$ is a weight matrix, and $\mathcal{E}_i$ is a set of hyperedges connected to node $v_i$. Then, the attention coefficient $\beta_{ij}$ for hyperedge representation $e^{l}_{j}$ is computed as follows:

\begin{equation}
\begin{aligned}
    \beta_{ij} &= \frac{exp \left( a^{T}_{2} r_{j} \right)}{\sum_{v_{i} \in e_{j}, e_{j}  \in \mathcal{E}_i} exp \left( a^{T}_{2} r_{i}\right)},\\
    r_{j} &= LeakyReLU \left( [ \mathrm{W_{2}} e^{l}_{j} \oplus  \mathrm{W_{1}} v^{l-1}_{i}  ] \right),
\end{aligned}
\end{equation}
where $a_{2} \in \mathbb{R}^{|\mathcal{V}| \times |\mathcal{E}|}$ is another context vector adopted to measure the importance of the hyperedges when aggregating the messages to the node representation.

\subsubsection{News (Node) Classification} The node representation $v^{\mathit{L}} \in  \mathbb{R}^{|\mathcal{V}| \times \mathit{d}}$ is a high-level representation resulted from node-level aggregation, where $\mathit{L}$ is the last layer of the hypergraph attention network. For the news classification layer, we use a fully connected layer with a trainable weight matrix $\mathrm{W_3}\in\mathbb{R}^{ \mathit{d} \times 2} $ and bias term $b$ to classify a node into two labels. To this end, we use Negative Log Likelihood to optimize the loss between the predicted logit of the $\textit{i}$-th node $\hat{y}_{i}$ and its ground-truth label, $y_{i}$ where $\mathcal{Y}_{train}$ is labels for training data.

\begin{equation}
\begin{aligned}
    \hat{y}_i  = Softmax(\:\mathit{f}(\mathrm{W_3} v^{\mathit{L}}_i + b) ),\\
    \mathcal{L} = \minus\sum_{y_i \in \mathcal{Y}_{train}}{\left( y_{i} log(\hat{y}_i) +  (1-y_i)log{(1-\hat{y}_i)} \right)}
\end{aligned}
\end{equation}

Note that the proposed framework works in the semi-supervised setting for transductive node classification in the hypergraph. This means the incidence matrix and features of all nodes in the hypergraph are accessible for training. However, there is no label leakage issues, since the labels of other nodes apart from the training data are unknown during the training phase~\cite{zhou2003learning}.


\begin{algorithm}
\caption{The overall process of HGFND}
\begin{algorithmic}[1]
\REQUIRE The news feature matrix $\mathcal{X}$,\\
\hspace*{1.2em} The propagation trees $\mathcal{P}$. 
\ENSURE The updated node representation on $\mathcal{V}$,\\
\hspace*{1.9em} The node-level attention weight $\alpha$,\\
\hspace*{1.9em} The hyperedge-level attention weight $\beta$.\\
\STATE Construct incidence matrix for the hypergraph $\mathcal{G}=(\mathcal{V}, \mathcal{E})$\\
$\mathbf{H} = \mathbf{H}^{user} \oplus \mathbf{H}^{time} \oplus \mathbf{H}^{entity}$;

\FOR{$P_i \in \mathcal{P}$}
\STATE Encode the propagation tree associated with news\\
$\overline{x}_i \leftarrow \textsc{Root}\left(GNN\left( P_i \right)\right)$;
\STATE Create node (news) representations for the hypergraph \\
${v}^0_{i} \leftarrow \textit{f} \left(\sigma \left( x_i \oplus \overline{x}_i \right)\right)$;

\ENDFOR
\FOR{$\mathit{l}=1 \dots L$}

\FOR{$e_j \in \mathcal{E}$}
\STATE Calculate hidden state $h_k$;
\STATE Calculate weight coefficient $\alpha_{jk}$;
\STATE Update the hyperedge representation $e^l_j$\\
$e^{l}_{j} \leftarrow \sigma \left(  \sum_{v_{k} \in e_{j}} \alpha_{jk} \mathrm{W_{1}} v^{l-1}_{k} \right)$;
\ENDFOR

\FOR{$v_i \in \mathcal{V}$}
\STATE Calculate hidden state $r_k$;
\STATE Calculate weight coefficient $\beta_{ij}$;
\STATE Update the node (news) representation $v^l_i$\\
$v^{l}_{i} \leftarrow  \sigma \left(  \sum_{e_{j} \in \mathcal{E}_{i}} \beta_{ij} \mathrm{W_{2}} e^{l}_{j} \right)$;
\ENDFOR

\ENDFOR
\RETURN {} $\{v^ \mathit{L}_i\}^{\mathit{N}}_{i=1}$, $\alpha$, $\beta$.
\end{algorithmic}

\end{algorithm}

\section{Experiments}
In this section, we aim to answer the following  questions:

\begin{itemize}
    \item \textbf{RQ1}: Can our proposed approach using hypergraph neural network improve fake news detection performance?
    \item \textbf{RQ2}: Does the proposed approach have a greater effect when only a limited number of training labels is given?
    \item \textbf{RQ3}: Is it more effective to use the hypergraph to represent news relations over a normal graph?
    \item \textbf{RQ4}: What types and combinations of hyperedges contribute most to performance of fake news detection?
    \item \textbf{RQ5}: Would the proposed method be useful for detecting users who tend to spread fake and true news?
    \end{itemize}  

\subsubsection{Infrastructures and Implementations}
All experiments are performed on a Linux machine with a GeForce RTX 6000 GPU (24 GB of VRAM), 64 GB of system memory, and Intel i7 @ 3.60GHz processors. Our models are implemented using PyTorch 1.9.0~\cite{paszke2019pytorch} and PytorchGemometric 2.0.2~\cite{fey2019fast}

\label{experiment_setup}
\subsubsection{Datasets}
We use FakeNewsNet~\cite{shu2020fakenewsnet} which contains news from two fact-checking platforms (Politifact, Gossipcop) with user's tweet and retweet associated to each news piece. However, since many Twitter accounts in FakeNewsNet are no longer accessible, we use the news and user information collected in UPFD~\cite{dou2021user}. In addition, UPFD provides news and user information in a form of pre-computed features with two different options: (1) news content, user's recent 200 tweets, and user's comments, which are encoded either in BERT~\cite{devlin2018bert} or word2vec~\cite{honnibal2017spacy}; and (2) user's 10 profile attributes such as the number of followers, timestamp, status verification, and so on. The full list of 10 profile features can be found in~\cite{han2020graph}. UPFD randomly splits data to train-val-test (20\%-10\%-70\%).

\begin{table}
\centering
\caption{Statistics of news propagation trees in UPFD~\cite{dou2021user}.}
\label{graph_statistics}
\begin{tabular}{ccccccc}
\hline
\textbf{Dataset} & \textbf{\# Graph} & \textbf{\# Fake}& \textbf{\# True}   & \textbf{\# Node } & \textbf{\# Edge} \\
\hline
\hline
Politifact & 314 & 157& 157 & 41,054 & 40,740\\
Gossipcop & 5,464 & 2,732& 2,732 & 314,262 & 308,798 \\
\hline
\end{tabular}
\end{table}


\subsubsection{Baselines}
We train the baselines with 10 different random seeds and the news samples provided in UPFD. Then, we classify the test data using the parameters of the model that has shown the best accuracy for validation. We evaluate the following representative baselines in fake news detection based on  (1) a propagation tree~\cite{bian2020rumor, monti2019fake}~\cite{han2020graph, dou2021user, shi2020masked, song2021temporally},   (2) a heterogeneous graph~\cite{ren2021fake}, and (3) a hypergraph~\cite{feng2019hypergraph} for a comprehensive evaluation. The features for news and users are based on the implementation of each baseline as follows:
\begin{itemize}
     \item \textbf{GNN-CL}~\cite{han2020graph}: We use DiffPool~\cite{ying2018hierarchical} to model structural information on a propagation tree based on user's profile attributes associated to each tweet/retweet. Due to the difference in the problem setting, training GNN with continuous learning is not considered in this study. 
    
    
     \item \textbf{BiGCN}~\cite{bian2020rumor}: It uses two GCN~\cite{kipf2016semi} modules in both top-down and bottom-up directions to capture propagation and dispersion patterns of news on a social network. User profile attributes are used to represent each tweet/retweet.
         
    \item \textbf{TGNF}~\cite{song2021temporally}: We use TGAT~\cite{xu2020inductive} to model temporal evolving pattern of a propagation tree by aggregating neighbors of tweet/retweet shared in proximity in time. It uses BERT to obtain the word vectors for tweet/retweet.

    \item \textbf{GTN}~\cite{matsumoto2021propagation}: We use Graph Transformer Network~\cite{shi2020masked} to encode a propagation tree, where the weight of an edge is determined by the time difference between tweet and retweet. Each tweet/retweet is represented by the corresponding user's recent 200 tweets, encoded by BERT.

     \item \textbf{HGAT}~\cite{ren2021fake}: We use a heterogeneous graph with two types of nodes (user and news) by linking them together if a user shares news. Based on two level aggregations with semantic and node level attention mechanism, it learns the relations between user and news nodes. We use BERT to encode each user's 200 recent tweets and news contents.
     
     \item \textbf{HGNN}~\cite{feng2019hypergraph}: It is a transductive node classification model using a hypergraph in a semi-supervised setting. It constructs the hypergraph based on the top 10 closest nodes according to euclidean distance for each node. Each node in the hypergraph is news content encoded by word2vec.
     
     \item \textbf{GCNFN}~\cite{monti2019fake}: It uses two-layer GCN~\cite{kipf2016semi} to model a propagation tree with the fusion of heterogeneous data such as user's profile, user's comments, and news contents which are encoded by word2vec.
     
     \item \textbf{UPFD}~\cite{dou2021user}: It combines various signals such as news text features and user preference features by jointly modeling news content and a propagation tree using a skip-connection. We experiment the model with three variants using GCN~\cite{kipf2016semi}, GAT~\cite{velivckovic2017graph}, and SAGE~\cite{hamilton2017inductive} as the GNN module. We use BERT to represent news content and the user preference based on the user's recent 200 tweets.

     
     

\end{itemize}

\subsubsection{Hyperparameters} We use unified size of the hidden dimension (128), batch size (128), optimizer (Adam), learning rate (0.001), dropout (0.3), and epoch (200) with early-stopping strategy. For the other parameters in the baselines, we apply the parameters suggested in each baseline. We use the pre-computed features of news and user's recent 200 tweet/retweet encoded by BERT~\cite{devlin2018bert} for the proposed model, where the initial feature dimension is 768, and we fix the number of layers in the hypergraph attention network to 2.

\subsection{RQ1: Performance on Fake News Detection}

\begin{table}
\centering
\caption{The results indicate mean ± standard deviation. Our proposed model outperforms all the baseline while meeting statistical significance under t-tests (p $<$ 0.05). OOM denotes out of memory, and the best performance is highlighted in boldface.}\label{result_full_tab}

\begin{tabular}{lcccc}
\hline
\multicolumn{1}{c}{\multirow{2}{*}{\textbf{Model}}}                        & \multicolumn{2}{c}{\textbf{Politifact}}                        & \multicolumn{2}{c}{\textbf{Gossipcop}}                         \\ \cmidrule(l){2-3} \cmidrule(l){4-5} 
\multicolumn{1}{c}{}                        &
\multicolumn{1}{c}{Accuracy} & \multicolumn{1}{c}{F1-score} & \multicolumn{1}{c}{Accuracy} & \multicolumn{1}{c}{F1-score} \\
\hline
\hline

GNN-CL& 65.79±8.96 & 65.02±9.46 &94.98±0.80& 94.94±0.80\\
BiGCN & 74.16±3.57 & 74.16±3.57 & 88.04±0.48&87.95±0.49\\
TGNF & 74.28±1.74 & 74.09±1.81& 85.07±0.08& 85.07±0.08\\
GTN  & 81.67±4.16& 81.53±4.35& 92.41±0.98&92.38±0.98\\
HGAT  & 81.53±1.16&80.47±1.75 & OOM & OOM\\

HGNN & 79.96±4.89 & 79.28±5.16 & 93.38±0.49 & 93.38±0.49\\
GCNFN & 80.63±4.23& 80.31±4.57 & 95.37±0.21&95.33±0.21\\
UPFD-GCN& 80.27±4.35& 80.16±4.41& 95.55±0.63&95.51±0.64\\
UPFD-GAT & 79.09±3.73&78.95±3.79 & 96.03±0.62&96.00±0.62\\
UPFD-SAGE & 80.40±4.22 & 80.13±4.65 & 96.38±0.48 & 96.36±0.48\\
%

\textbf{HGFND}&\textbf{91.11±1.89} & \textbf{91.11±1.89} & \textbf{97.46±0.30} & \textbf{97.46±0.30}\\
\hline
\end{tabular}
\end{table}

Table~\ref{result_full_tab} shows the performance of all compared methods. HGFND achieves the best performance across all datasets w.r.t. both accuracy and F1-score. Based on the evaluation, we make three primary observations as follows:

\begin{table*}[!htbp]
\centering
\caption{Test accuracy by varying the percentage of available labels for training data where OOM denotes Out of memory.}\label{result_downsample}

\begin{tabular}{lcccccccccc}
\hline
\multicolumn{1}{c}{\multirow{2}{*}{\textbf{Model}}}    & \multicolumn{4}{c}{\textbf{Politifact} (small dataset)}          & \multicolumn{6}{c}{\textbf{Gossipcop} (large dataset)}
              \\ \cmidrule(l){2-5} \cmidrule(l){6-11}
              & 100\% & 75\%        & 50\%        & 25\%        & 100\% & 75\% & 50\%        & 25\%        & 5\%& 2.5\%        \\
\hline
\hline

\textbf{HGFND}     &   \textbf{91.13±1.8}  & \textbf{90.89±2.6} & \textbf{87.91±3.0} & \textbf{85.61±6.0} &  \textbf{97.46±0.3}  & \textbf{97.32±0.3} & \textbf{97.13±0.4} & \textbf{97.05±0.3} & \textbf{96.69±0.4}& \textbf{96.45±0.4} \\
UPFD-SAGE     &  80.40±4.2   &    77.62±2.8       &      75.87±2.7      &     68.24±5.4      &  96.38±0.4 & 96.26±0.6  &     95.87±0.4      &       94.65±0.4    &   92.42±1.4  & 87.81±3.1     \\
UPFD-GAT      &   79.09±3.7  &    77.24±3.2       &     75.51±4.0      &     69.29±6.1      &  96.03±0.6  & 95.46±0.6 &     95.05±0.4      &     93.46±0.2      &    90.21±2.9  & 84.99±3.4    \\
UPFD-GCN      &  80.21±4.3   &     76.86±5.3      &     76.70±5.3      &     70.25±4.0      &   95.55±0.6& 95.38±0.4  &     92.93±3.1      &     89.22±3.2      &    85.77±2.3  & 83.14±3.6    \\
GCNFN         &  80.63±4.2   &     78.65±6.1      &     77.16±6.2      &     71.53±8.1      &   95.37±0.2 & 95.11±0.3 &      94.65±0.6      &      91.19±1.7     &    89.75±2.8 & 88.82±3.2      \\

HGNN          &  79.96±4.8   &     79.45±4.2     &      78.69±4.6      &    75.71±4.5        &  93.38±0.4 & 93.34±0.4 &     93.21±0.4      &      93.11±0.4     &     92.56±0.5 & 91.37±1.6    \\
HGAT          &  81.53±1.1   &     80.00±2.0      &     77.51±3.6      &     69.04±5.7      &   OOM & OOM &     80.95±7.6      &     75.73±6.0      &     68.06±5.2    &60.12±3.9 \\
GTN           &   81.67±4.1  &     78.85±3.4      &     69.21±5.4      &     64.28±5.0      &  92.41±0.9& 89.54±2.3   &      87.38±3.4     &     83.81±3.9      &     81.34±4.4   & 75.82±6.8  \\
    TGNF          &  74.28±1.7   &     70.30±1.8      &      63.94±7.8     &      59.46±1.9     &  85.07±0.0& 84.53±0.4   &      83.54±1.7     &     78.31±2.2      &    65.54±1.4   & 65.29±0.8   \\
BiGCN         &   74.16±3.5  &     74.02±4.0      &     72.44±5.5      &      68.20±8.4     &  88.04±0.4& 84.54±0.6   &      83.99±0.7     &     82.13±0.9      &    79.54±0.6   & 78.83±1.2   \\

GNN-CL        &  65.79±8.9   &     64.23±6.0      &     57.34±9.0      &      56.75±5.9     &   94.98±0.8& 93.73±0.4  &     91.98±0.8      &     91.62±1.5      &    84.42±1.8 &80.36±2.8     \\
\hline
\end{tabular}
\end{table*}

First, in contrast to propagation-based methods, HGFND constantly shows better performance. While variations of UPFD show comparable performance in Gossipcop, they show much lower performance in Politifact compared to HGFND. Considering that the size of Gossipcop is $\sim$17 times larger than Politifact, the difference in performances on two datasets shows the robustness of HGFND to available training labels.

Second, in contrast to the heterogeneous graph-based method, HGFND has advantages in terms of space usage and expressive power in capturing relational information. HGAT runs out of memory since it includes all users in the dataset as nodes and connects them to a news node, while HGFND includes users when they share more than one news. HGAT is also limited in capturing the relational information between news nodes, as its attention mechanisms only aggregates the the local neighbors (i.e., user nodes) of a single news node.

Third, in comparison with the hypergraph-based method, HGFND shows greater performance than HGNN by $\sim$12\% on Politifact and $\sim$4\% on Gossipcop w.r.t both accuracy and F1-score. This result proves that relying on the semantic similarity between nodes can involve noisy and task-irrelevant information to fake news detection and finding meaningful relation for news is important for the quality of the hypergraph.

\subsection{RQ2: Is HGFND more effective with limited train labels?}

To answer RQ2, we train and evaluate the models with a subset of training labels. To this end, we vary the percentage of training labels according to the two different sizes of the datasets; Gossipcop contains more labeled data than Politifact. When randomly sampling training data, we remove the unselected training data from the hypergraph, and also remove the corresponding propagation tree from the dataset.

As shown in Table.~\ref{result_downsample}, we find consistent improvements over the baselines regarding accuracy under both limited and sufficient data conditions. For a clear comparison, we compare the highest and lowest accuracy of the proposed model per dataset. In Politifact, HGFND's accuracy decreases by $\sim$6\% (91.13 $\rightarrow$ 85.61). In Gossipcop, HGFND's accuracy decreases by $\sim$1\% (97.46 $\rightarrow$ 96.45). While other baselines suffer a significant decrement, HGFND shows robust performance to the limited number of training labels.

Notably, another hypergraph-based method, HGNN~\cite{feng2019hypergraph}, also shows robust performance to the size of training labels than the other baselines. HGNN's accuracy decreases by $\sim$5\% (79.96 $\rightarrow$ 75.71) in Politifact, and $\sim$2\% (93.38 $\rightarrow$ 91.37) in Gossipcop. This experimental result supports that leveraging a hypergraph can help to better utilize training labels by learning and preserving the group-wise interaction among news.

\subsection{RQ3: Advantage of a Hypergraph over a Normal Graph}

To answer RQ3, we compare the performance between a series of GNN models, and the proposed methods based on the hypergraph constructed in Section~\ref{Hypergraph_Construction}. As it is incompatible to directly apply GNNs to a hypergraph, we first convert the hypergraph into graph structure using clique expansion~\cite{zhou2006learning}, which replaces hyperedges with cliques to obtain a graph. In this experiment, GNNs define node features as news content encoded by BERT, and classify the unlabeled nodes in a semi-supervised setting for transductive node classification. As compared in Table~\ref{result_graph_hypergraph}, GCN~\cite{kipf2016semi}, GAT~\cite{velivckovic2017graph}, and SAGE~\cite{hamilton2017inductive} constantly show worse performance than HGFND across all datasets w.r.t both accuracy and F1-score. The performance gap can be mainly attributed to undesired information loss that occurs when clique expansion converts group-wise interactions in the hypergraph to pairwise relations~\cite{hein2013total}.

\begin{table}[!htbp]
\centering
\caption{Comparison of performance between the proposed model and a series of graph neural networks based on clique expansion.}\label{result_graph_hypergraph}
\begin{tabular}{lcccc}
\hline
\multicolumn{1}{c}{\multirow{2}{*}{\textbf{Model}}} & \multicolumn{2}{c}{\textbf{Politifact}}                        & \multicolumn{2}{c}{\textbf{Gossipcop}}                         \\ \cmidrule(l){2-3} \cmidrule(l){4-5} 
\multicolumn{1}{c}{}                        & \multicolumn{1}{c}{Accuracy} & \multicolumn{1}{c}{F1-score} & \multicolumn{1}{c}{Accuracy} & \multicolumn{1}{c}{F1-score} \\
\hline
\hline
\textbf{HGFND}& \textbf{91.13±1.89} & \textbf{91.11±1.89}& \textbf{97.46±0.30} & \textbf{97.46±0.30}\\
SAGE~\cite{hamilton2017inductive}             & 80.84±3.55 & 80.80±3.59& 94.79±0.38 & 94.79±0.38\\
GAT~\cite{velivckovic2017graph}             & 80.09±4.49 & 79.98±4.62& 94.75±0.37 & 94.75±0.37\\
GCN~\cite{kipf2016semi}             & 79.90±4.07 & 79.77±4.20&94.69±0.41 & 94.68±0.42 \\
\hline
\end{tabular}
\end{table}


\subsection{RQ4: Contribution of Hyperedge Types on Performance}
To answer RQ4, we conduct an ablation study on HGFND with combinations of the three different types of hyperedges (\textit{User},  \textit{Time}, and  \textit{Entity}). As shown in Table~\ref{result_edges_tab}, combining all types of hyperedges yields the best performance. Each hyperedge type is complementary to the other, as performance always increases when adding more types of hyperedge. When comparing the contribution of each hyperedge type, \textit{User} hyperedge shows the highest contribution to the performance on both datasets in every combination with other hyperedge types. \textit{Time} and \textit{Entity} hyperedges appear to be the second most contributing type of hyperedges, but the contribution is different for each dataset.


Overall, the experimental results support that users play a key role in detecting fake news. Based on such observation, we ask the following question, \textit{``How does users' news sharing behavior help to detect fake news?''}. We answer it next.

\begin{table}
\centering
\caption{Evaluation of the proposed model (HGFND) on combination of different types of hyperedges. U, T, and E symbols indicate the hyperedge types of user, time, and entity respectively.}\label{result_edges_tab}

\begin{tabular}{lcccc}
\hline
\multicolumn{1}{c}{\textbf{Type}} & \multicolumn{2}{c}{\textbf{Politifact}}                        & \multicolumn{2}{c}{\textbf{Gossipcop}} \\ \cmidrule(l){1-1} \cmidrule(l){2-3} \cmidrule(l){4-5} 
\multicolumn{1}{c}{U T E}                        & \multicolumn{1}{c}{Accuracy} & \multicolumn{1}{c}{F1-score} & \multicolumn{1}{c}{Accuracy} & \multicolumn{1}{c}{F1-score} \\
\hline
\hline
\:\:\checkmark\:\checkmark\:\checkmark& \textbf{91.13±1.89} & \textbf{91.11±1.89}& \textbf{97.46±0.30} & \textbf{97.46±0.30} \\
\:\:\checkmark\:\checkmark\:\:-& 89.72±3.11 & 89.70±3.14& 97.38±0.25 & 97.38±0.25\\
\:\:\checkmark\:\:-\:\:\checkmark&89.90±2.52 & 89.89±2.52& 97.41±0.28& 97.41±0.28\\
\:\:\:-\:\:\checkmark\:\checkmark& 85.24±4.18 & 85.13±4.30 & 90.01±2.57 & 90.00±2.58\\
\:\:\checkmark\:\:-\:\:\:-\:\:\:& 87.23±2.32 & 87.17±2.34 & 97.29±0.16 & 97.29±0.16\\
\:\:\:-\:\:\checkmark\:\:-& 83.61±3.28 & 83.49±3.38 & 72.15±3.26 & 71.85±3.39\\
\:\:\:-\:\:\:-\:\:\checkmark & 76.28±3.09 & 76.15±3.10 & 84.46±0.20 & 84.42±0.20\\
\hline
\end{tabular}
\end{table}

\begin{table}
\centering
\caption{Statistics of the proposed news hypergraph in Section~\ref{Hypergraph_Construction}. Node degree is the number of hyperedges connected to a node. Hyperedge size is the number of nodes included in a hyperedge.}
\label{hypergraph_statistics}

\begin{tabular}{C{2.4cm}C{0.6cm}C{0.6cm}C{0.6cm}C{0.6cm}C{0.6cm}C{0.6cm}}
\hline
\multicolumn{1}{c}{\multirow{2}{*}{\textbf{Measure}}}    & \multicolumn{3}{c}{\textbf{Politifact}}          & \multicolumn{3}{c}{\textbf{Gossipcop}}
              \\ \cmidrule(l){2-4} \cmidrule(l){5-7}
              & U & T & E         & U & T & E        \\
\hline
\hline

\# hyperedges     &  2,953  & 1,717 & 1,040 & 15,646  & 14,586 & 12,673  \\
Avg. hyperedge size     &  2.56  & 3.52 & 4.88  &  13.19  & 6.23 & 6.84 \\
Max. hyperedge size     &  21  & 13 & 80  &  2,426  & 25 & 992  \\
Avg. node degree    &  25.59  & 19.35 & 17.94 &  37.94  & 16.64 & 16.31 \\
Max. node degree    &  153  & 186 & 147  &  156  & 124 & 471 \\

\hline

\end{tabular}
\end{table}
\subsection{RQ5: Interpreting Users' News Sharing Behavior}

\begin{figure}
\centering
\includegraphics[width=0.5\textwidth]{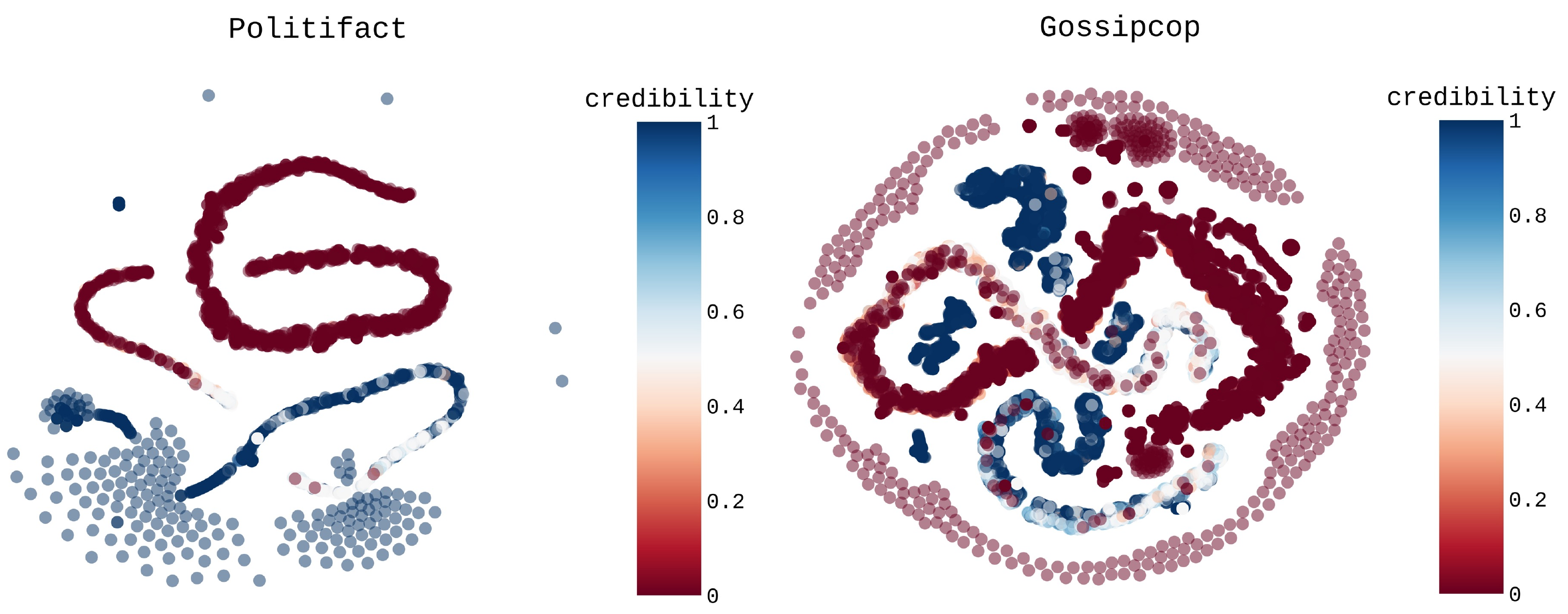}
\caption{t-SNE plot on the representations of the \textit{User} hyperedge in 2D space. The credibility score (the higher, the credible) is shown by color and value.}
\label{user_embed}
\end{figure}

To answer RQ5, we examine whether HGFND can learn to characterize users who tend to spread fake or true news. To this end, we measure the credibility of a user, defined as the percentage of shared true news, as follows:

\begin{equation}
\label{credibility_user_eq}
\textit{credibility}_{i} = \frac{ \sum_{y_j \in \mathcal{Y}^{user}_i} {y_j}}{|\mathcal{E}^{user}_i|}\:,\:\mathrm{where}\:\:\mathcal{E}^{user}_{i} \subset \mathcal{E}^{user}
\end{equation}
where $\textit{credibility}_i$ is the credibility score of \textit{i}-th user, $\mathcal{E}^{user}$ is a set of \textit{User} hyperedges, and $|\mathcal{E}^{user}_i|$ is the number of nodes connected to $\mathcal{E}^{user}_i$. Here, $\mathcal{Y}^{user}_i$ is a set of labels for nodes connected to $\mathcal{E}^{user}_i$ where the label is 0 if fake and 1 if true.

To visualize the representation of $\mathcal{E}^{user}$, we first apply t-SNE~\cite{pedregosa2011scikit} and calculate the credibility score assigned to users. As shown in Fig.~\ref{user_embed}, HGFND separates the \textit{User} hyperedges into two distinctive groups with high and low credibility scores, respectively.

\begin{figure}[H]
\centering
\includegraphics[width=0.5\textwidth]{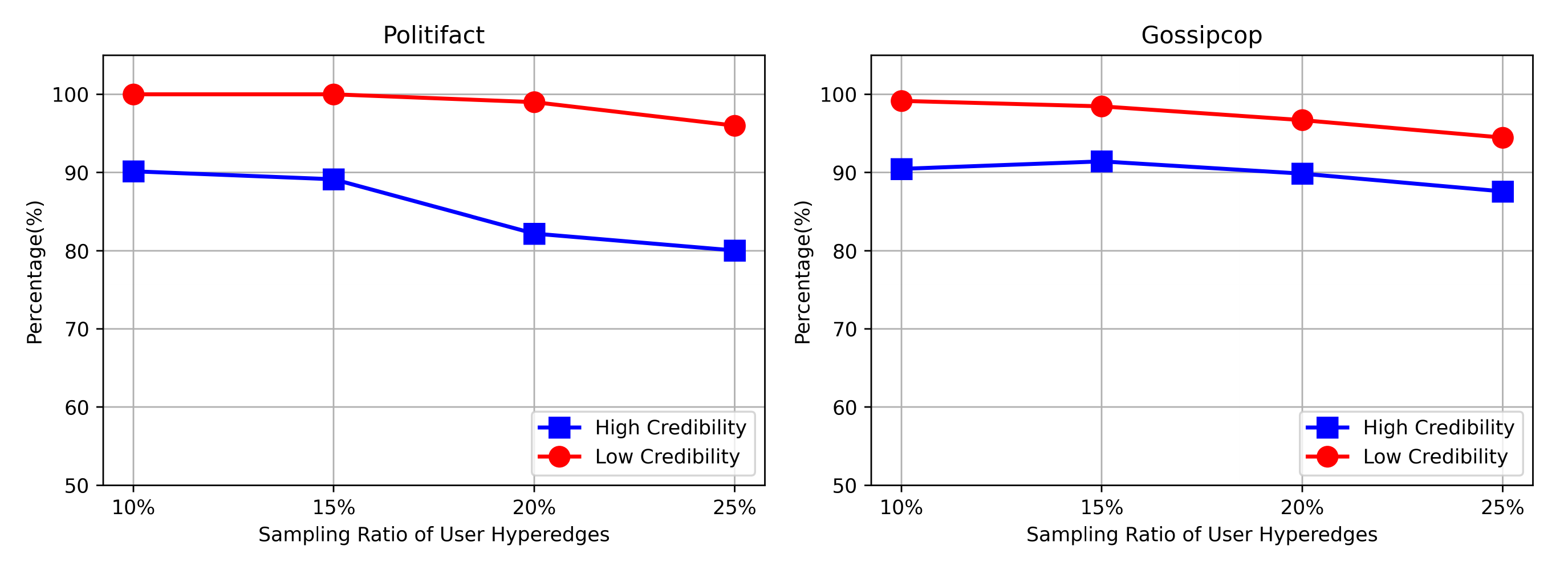}
\caption{The percentage of users with high and low credibility that are sampled from \textit{User} hyperedges. Users are selected based on the sampling ratio (y-axis) according to the top/bottom attention scores of the hyperedges.}
\label{user_detect}
\end{figure}

We further answer whether the attention scores assigned to \textit{User} hyperedges can help to identify users who tend to spread true or fake news. To this end, we follow two sampling steps: (1) sampling users assigned with  high and low attention scores, respectively, and (2) checking how many users are classified to have high or low credibility. Here, a user's credibility is high if the credibility score exceeds 0.9, and low if the credibility score is less than 0.1. For example, we sample the top 5\% of users with the highest attention scores and check if their credibility scores are above or below the thresholds. 

Fig.~\ref{user_detect} shows the percentage of users with high and low credibility from selected users with varying sampling ratio (10\%, 15\%, 20\%, and 25\%) based on the top/bottom attention scores. For both datasets, the percentage of users with high credibility lies between 80\% and 95\%, and the percentage of users with low credibility is between 95\% and 100\%. Based on the results, we show that the proposed model can implicitly learn the credibility of users based on the attention mechanism. This finding opens a new avenue to interpret the model's decision through the lens of users' news sharing behavior.


\section{Conclusion}
In this study, we propose a novel hypergraph-based neural network solving the problem of relational fake news detection with limited labeled news data. Our approach can better utilize the relational information between labeled and unlabeled data by capturing group-wise interactions among news pieces. Based on two real-world benchmark datasets for fake news detection, we show the superiority of the proposed model over the state-of-the-art methods on both normal and limited data settings. Experimental results demonstrate that focusing on useful subsets of the relations is important, and identifying susceptible users plays a key role in fake news detection.

\section{Limitations \& Future Work }
The results here are not without limitations. Some public datasets~\cite{ma2016detecting, ma2017detect, kochkina2017turing} are not considered in this study as they are not in line with the definition of fake news, which is an intentional and verifiable false news published by a news outlet. Furthermore, the type information of each hyperedge is not considered in the framework, which can potentially misguide the inference when more types of hyperedges are added. Additionally, we round timestamp for tweet/retweet differently for the two datasets (i.e. Politifact in days and Gossipcop in hours). The rounding method is empirically opted according to the criteria that yields the best performance.


Future work will focus on finding more types of news relations, such as news sharing similar event schema, exploitative narratives, and geolocation. Based on these comprehensive news relations, we will employ a heterogeneous hypergraph to better reflect different contribution per the type of hyperedge. Lastly, we aim to dynamically learn and exploit the structure of the hypergraph by merging and splitting existing hyperedges. The structure information will help to detect fake news when only a few news relations are 
known.

\section{Reproducibility \& Social Impact}
Existing fake news-related studies show difficulties in benchmarking their datasets, as user information is in many times not publicly available due to data sharing policies. To prevent this problem, we use pre-computed features for news and users provided by UPFD~\cite{dou2021user}. Through this, the proposed method can reproduce the result at any time without additional data collection procedure. Furthermore, as news and user features are encoded, it is expected to be safer in terms of information security. The code and data are publicly available at \url{https://github.com/ujeong1/IEEEBigdata22_HGFND}.

We acknowledge that our work provides only preliminary solutions, and it is far from objectively discerning the authenticity of news in a real-world scenario due to the complexity and diversity of fake news. Furthermore, the result of the proposed approach can be biased due to the existing bias in the dataset or the algorithm. Nevertheless, with our work, we hope to open a potential research directions in the problem of relational fake news detection while broadening discussions in mitigating the spread of fake news on online social networks.

\section{Acknowledgements}
This work was supported by the Office of Naval Research under Award No. N00014-21-1-4002. Opinions, interpretations, conclusions, and recommendations
are those of the authors.

\bibliographystyle{IEEEtran}
\bibliography{references}

\end{document}